\documentclass[preprint,aps,floatfix,a4paper]{revtex4-2}

\usepackage{graphicx}
\usepackage{hyperref}
\usepackage{amsmath}
\usepackage{amssymb}
\usepackage{amsthm}
\usepackage{braket}
\usepackage{dsfont}
\usepackage{color}

\begin{document}

\title{Boundary effect and correlations in fermionic Gaussian states}

\author{{Jinhyeok} {Ryu}}
\email{jisungin28@gmail.com}
\author{{Jaeyoon} {Cho}}
\email{choooir@gmail.com}
\affiliation{{Department of Physics and Research Institute of Natural Science}, {Gyeongsang National University}, {{Jinju}, {52828}, {Korea}}}

\begin{abstract}
The effect of boundaries on the bulk properties of quantum many-body systems is an intriguing subject of study. One can define a boundary effect function, which quantifies the change in the ground state as a function of the distance from the boundary. This function serves as an upper bound for the correlation functions and the entanglement entropies in the thermodynamic limit. Here, we perform numerical analyses of the boundary effect function for one-dimensional free-fermion models. We find that the upper bound established by the boundary effect fuction is tight for the examined systems, providing a deep insight into how correlations and entanglement are developed in the ground state as the system size grows. As a by-product, we derive a general fidelity formula for fermionic Gaussian states in a self-contained manner, rendering the formula easier to apprehend.
\end{abstract}

\maketitle
\newpage

\section{Introduction}

Quantum many-body theory primarily aims to understand phenomena in systems of infinite size, namely, the thermodynamic limit.
In addition, the influences of boundaries also constitute essential subjects of interest.
It is folklore that when a finite system is sufficiently large, the bulk---the central region far from the boundary---exhibits the thermodynamic properties while the region near the boundary is altered, reflecting the bulk properties.
A quintessential example is the concept of bulk-boundary correspondence: topological orders in the bulk are manifested as the boundary modes~\cite{wen91,hat93,lu12}.
That the boundary effect depends on the bulk properties is enlightening and conceptually reasonable.
Yet, there remain many aspects to be clarified in this picture.
For example, where is the border between the bulk and the boundary? 
If the border is obscure, how are the two distinct features of the bulk and the boundary interpolated?
How can such interpolation be characterized and quantified?

The reasoning on these questions leads us to an insight into how correlations are established in many-body ground states.
To illustrate the idea, consider a certain sequence of $n$-spin local Hamiltonians $H^{(n)}$ with increasing $n$, and let $\ket{\Psi_0^{(n)}}$ be the ground state of $H^{(n)}$.
Typically, the different-sized Hamiltonians are constructed with a particular rule.
For example, Heisenberg chains are defined in terms of the interaction term, thus the Hamiltonian is consistently determined for any system size.
The focus here is on how $\ket{\Psi_0^{(n)}}$ morphs into $\ket{\Psi_0^{(n+1)}}$ when the system size increases.

For the moment, let us make a radical assumption that $\ket{\Psi_0^{(n)}}$ is transformed to $\ket{\Psi_0^{(n+1)}}$ by altering only a finite region near the $(n+1)$-th spin at the boundary.
In this case, the ground state cannot develop long-range correlations as the unaltered bulk region away from the boundary is completely uncorrelated with the newly created outer region.
In other words, that unaltered region does not \emph{recognize} the change of the system size, which means that the region remains the same even in the thermodynamic limit.
One can also relate this situation to the famous open problem concerning the scaling of entanglement in the ground state~\cite{eis10}.
When the above assumption is satisfied, the ground state obeys the entanglement area law~\cite{cho14,cho15}.

The above scenario is, of course, unrealistic.
In reality, the influence made by the change at the boundary gradually decreases in the direction toward the bulk.
The nature of this decrease would then impact on the correlation and entanglement in the ground state, and also characterize the convergence to the thermodynamic limit, as exemplified above.
This concept was carefully materialized in Refs.~\cite{cho14,cho15} by one of the present authors.
The key quantity is the boundary effect function (BEF).
For simplicity, consider a one-dimensional array of spins and suppose that the state $\ket{\Psi_0^{(n)}}$ is extended to $\ket{\Psi_0^{(n+1)}}$ by adding the $(n+1)$-th spin to the right.
Let the reduced density matrices of these states after removing $r+1$ rightmost spins, i.e., the density matrices of the spins at the sites from $1$ to $n-r$, be $\Omega_{1,n-r}^{(n)}$ and $\Omega_{1,n-r}^{(n+1)}$, respectively.
Denoting by $F(\rho,\sigma)\equiv \text{Tr}\sqrt{\sqrt{\rho}\sigma\sqrt{\rho}}$ the fidelity between two density matrices $\rho$ and $\sigma$, we define the BEF as
\begin{equation}
    \mu_n (r) = \sqrt{1 - F\left(\Omega_{1,n-r}^{(n)},\Omega_{1,n-r}^{(n+1)}\right)}.
    \label{eq:bef}
\end{equation}
Intuitively, this quantifies how well the region away from the boundary can distinguish between different system sizes.
The BEF $\mu_{\infty}(r)$ characterizes the thermodynamic properties of the system.
The definition can be generalized straightforwardly for higher dimension in several ways.
For example, one way is to obtain the two density matrices in Eq.~\eqref{eq:bef} by removing those spins within distance $r$ from the entire boundary.

The BEF indeed restricts the correlation in the ground state.
If $\mu_\infty (r)$ decreases exponentially with $r$, the ground state exhibits local correlations in that any two-point correlation function decays exponentially with the distance, called the exponential clustering theorem~\cite{fre85,has06,nac06}, and the entanglement entropies exhibit an area-law scaling~\cite{has07,eis10,bra13,cho14,cho15,cho18}.
A naturally ensuing question is then concerning the \emph{tightness} of the upper bound provided by the BEF, i.e., whether the exponentially decaying BEF is also a \emph{necessary} condition for such local nature of correlations.
At first glance, it seems natural that BEFs decay exponentially when all correlation functions in the bulk do so.
However, this is highly nontrivial.
If it can be proven, it leads to a significant advance in our understanding of ground-state entanglement, partially solving one of the long-lasted open problems in quantum information and many-body theories~\cite{eis10}.

The aim of this paper is two-fold.
First, we numerically calculate BEFs for actual physical models of one-dimensional free fermions.
In the examined systems, we observe that the extent of the BEF is finite for gapped systems and divergent for gapless systems, following the behavior of correlation functions in the bulk.
This result thus supports the above-mentioned idea that the correlation properties in the bulk and boundary are strongly tied. 
For these analyses, it is necessary to calculate the fidelities between two fermionic Gaussian states.
Rather surprisingly, the explicit fidelity formula required for our investigation was presented only recently in Ref.~\cite{swi19} based on the general framework presented in Ref.~\cite{bra05}.
However, the formula is somewhat inaccessible as the wide generality of the underlying framework in the latter, slightly axiomatic in some aspects, significantly obscures the derivation logic in the former.
The second aim of this paper is thus to derive the fidelity for fermionic Gaussian states directly from scratch, thereby clarifying the derivation process and making the fidelity formula more accessible to a broad readership.

\section{Fermionic Gaussian states}

Consider a system of $n$ fermionic modes, described by $2n$ fermion operators $\{f_j\}$ and $\{f_j^\dagger\}$ with $j\in[1,n]$.
It is often more convenient to use $2n$ Majorana operators $c_{2j-1} = f_j + f_j^\dagger$ and $c_{2j} = -i (f_j - f_j^\dagger)$ without distinguishing between the creation and annihilation operators.
Fermionic Gaussian states are defined as those density matrices written as~\cite{bra05}
\begin{equation}
    \rho = \frac{1}{Z} e^{\frac{i}{4} \pmb{c}^T \pmb{H} \pmb{c}},
    \label{eq:thermal}
\end{equation}
where $\pmb{c} = (c_1 ~ c_2 ~ \cdots ~ c_{2n})^T$ is the column vector of the Majorana operators, $\pmb{H} = - \pmb{H}^T$ is a real skew-symmetric matrix, and $Z = \text{Tr} (e^{\frac{i}{4} \pmb{c}^T \pmb{H} \pmb{c}})$ is the normalization factor.
The matrix $\pmb{H}$ is decomposed as
\begin{equation}
    \pmb{H} = \pmb{R} \left[ \bigoplus_{j=1}^n \left(\begin{matrix}
        0 & \epsilon_j \\ -\epsilon_j & 0
    \end{matrix}\right)\right] \pmb{R}^T
    \label{eq:h}
\end{equation}
with real orthogonal matrix $\pmb{R}$ satisfying $\pmb{R}\pmb{R}^T = \pmb{I}$, where $\epsilon_j \ge 0$.
This decomposition allows us to make a basis transformation $\pmb{c} \rightarrow \pmb{\tilde{c}} = \pmb{R}^T \pmb{c}$, which preserves the anticommutativity $\{\tilde{c}_j, \tilde{c}_k\} \equiv \tilde{c}_j \tilde{c}_k + \tilde{c}_k \tilde{c}_j  = 2 \delta_{jk} I$.
In this basis, the state is written as
\begin{equation}
    \rho = \frac{1}{2^n} \prod_{j=1}^{n} ( I + i \lambda_j \tilde{c}_{2j-1} \tilde{c}_{2j}) = \prod_{j=1}^{n} \left(\frac{1-\lambda_j}{2} \tilde{f}_j \tilde{f}_j^\dagger + \frac{1+\lambda_j}{2} \tilde{f}_j^\dagger \tilde{f}_j \right) ,
    \label{eq:ground}
\end{equation}
where $\tilde{f}_j$ and $\tilde{f}_j^\dagger$ are the fermion operators in the transformed basis, and $\lambda_j \equiv \tanh \frac{\epsilon_j}{2}$.
Note that the ground state corresponds to the case $\epsilon_j \rightarrow \infty$, i.e., $\lambda_j = 1$, for all $j$.
While Eq.~\eqref{eq:thermal} is apparently suitable for representing thermal states of quadratic fermion Hamiltonians,
it is often awkward for handling ground states due to the diverging parameter $\epsilon_j$.
For our purpose, the representation in Eq.~\eqref{eq:ground} is thus more preferable.
However, this form lacks the convenience that the algebra of exponential functions provides.
In particular, we need to obtain fidelities of Gaussian states, for which products of density matrices should be worked out.
This difficulty is overcome elegantly by introducing the Grassmann algebra~\cite{bra05}.
The details are elaborated in the Appendix.

Using the Wick's theorem, the system is fully determined in every respect by two-point correlation functions $\Gamma_{jk} = \frac{i}{2} \text{Tr} ( \rho [c_j, c_k])$~\cite{gau60}.
Majorana operators have a convenient property that any product of different Majorana operators is traceless:
\begin{equation}
    \text{Tr} (c_{j_1} c_{j_2} \cdots c_{j_m}) = 0
    \label{eq:trace}
\end{equation}
for different $c_j$s with $m \ge 1$.
Using this, one can find from Eq.~\eqref{eq:ground} that the matrix $\Gamma_{jk}$ is identical to
\begin{equation}
    \pmb{\Gamma} = \pmb{R} \left[ \bigoplus_{j=1}^{n} 
    \left(\begin{matrix}
        0 & \lambda_j \\ - \lambda_j & 0
    \end{matrix}\right) \right] \pmb{R}^T,
    \label{eq:correlation}
\end{equation}
which is referred to as the correlation matrix.
Note that this matrix is represented in the original basis $\{c_j\}$.
For any subsystem containing only a subset of $\{c_j\}$, the corresponding submatrix of Eq.~\eqref{eq:correlation} can be taken.
Apparently, this submatrix contains all the two-point correlation functions $\Gamma_{jk}$ for the subsystem.
This implies that the extracted submatrix automatically becomes the correlation matrix for the subsystem.

The fidelity between two density matrices $\rho$ and $\sigma$ is defined as 
\begin{equation}
    F(\rho,\sigma) = \text{Tr} \sqrt{\sqrt{\rho} \sigma \sqrt{\rho}}.   
\end{equation}
When the Gaussian states are represented as in Eq.~\eqref{eq:thermal}, the fidelity can be obtained straightforwardly as explained in Ref.~\cite{ban14}.
However, the states represented as in Eq.~\eqref{eq:ground}, suitable for our work, should be treated differently.
For two fermionic Gaussian states characterized by correlation matrices $\pmb{\Gamma}_\rho$ and $\pmb{\Gamma}_\sigma$, the fidelity is given by
\begin{equation}
    F(\rho, \sigma)  = \det\left(
        \frac{\pmb{\Gamma}_\rho \pmb{\Gamma}_\sigma - \pmb{I}}{2}
    \right)^{1/4} \det\left(
        \pmb{I} + \sqrt{
            \pmb{I} + \left\{
                \left(
                    \pmb{\Gamma}_\rho + \pmb{\Gamma}_\sigma
                \right) \left(
                    \pmb{\Gamma}_\rho \pmb{\Gamma}_\sigma - \pmb{I}
                \right)^{-1}
            \right\}^2
        }
    \right)^{1/4}.
    \label{eq:fidelity}
\end{equation}
The details of the derivation is elaborated in the Appendix.
Note that Eq.~\eqref{eq:fidelity} contains a potentially singular term $(\pmb{\Gamma}_\rho \pmb{\Gamma}_\sigma - \pmb{I})^{-1}$.
As the correlation matrices have pure imaginary eigenvalues with the modulus not exceeding one, this term becomes singular only when $\pmb{\Gamma}_\rho$ and $\pmb{\Gamma}_\sigma$ have at least one common eigenvector with the respective eigenvalues $+i$ and $-i$, which means that one is the single fermion and the other is the vacuum for that common eigenmode.
In this case, the fidelity vanishes.
Consequently, one can first check if the first determinant in Eq.~\eqref{eq:fidelity} vanishes, for which $F(\rho,\sigma)=0$, and otherwise calculate the second one without the issue of the singularity.

\section{Boundary effect functions in free-fermion models}

\subsection{Transverse-field Ising model}

As one of the prominent one-dimensional free-fermion models, we consider the transverse-field Ising model described by Hamiltonian
\begin{equation}
    H = - \sum_{j=1}^{n-1} \sigma_j^x \sigma_{j+1}^x - h \sum_{j=1}^{n} \sigma_j^z,
    \label{eq:ising}
\end{equation}
where $\sigma_j^x$, $\sigma_j^y$, and $\sigma_j^z$ are the Pauli operators acting on the $j$-th spin.
Through the Jordan-Wigner transformation $f_j \equiv \sigma_j^- \prod_{k<j} \sigma_k^z$ with $\sigma_j^- = \frac{1}{2} (\sigma_j^x - i \sigma_j^y)$, the Hamiltonian is transformed to a free-fermion one:
\begin{equation}
    H = - \sum_{j=1}^{n-1} (f_j^\dagger - f_j) (f_{j+1}^\dagger + f_{j+1}) - 2h \sum_{j=1}^{n} f_j^\dagger f_j.
\end{equation}
This system undergoes a quantum phase transition at the critical point $h =1$.
Apart from this point, the ground state produces exponentially decaying correlation functions.

\begin{figure}
    \includegraphics[width=0.33\textwidth]{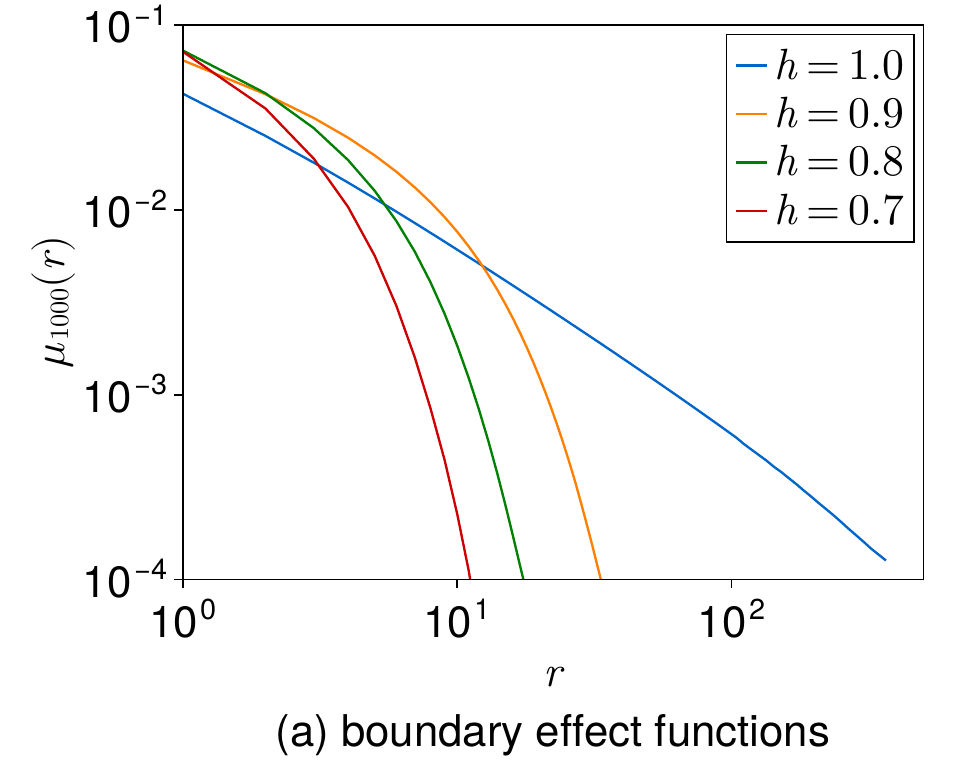}%
    \includegraphics[width=0.33\textwidth]{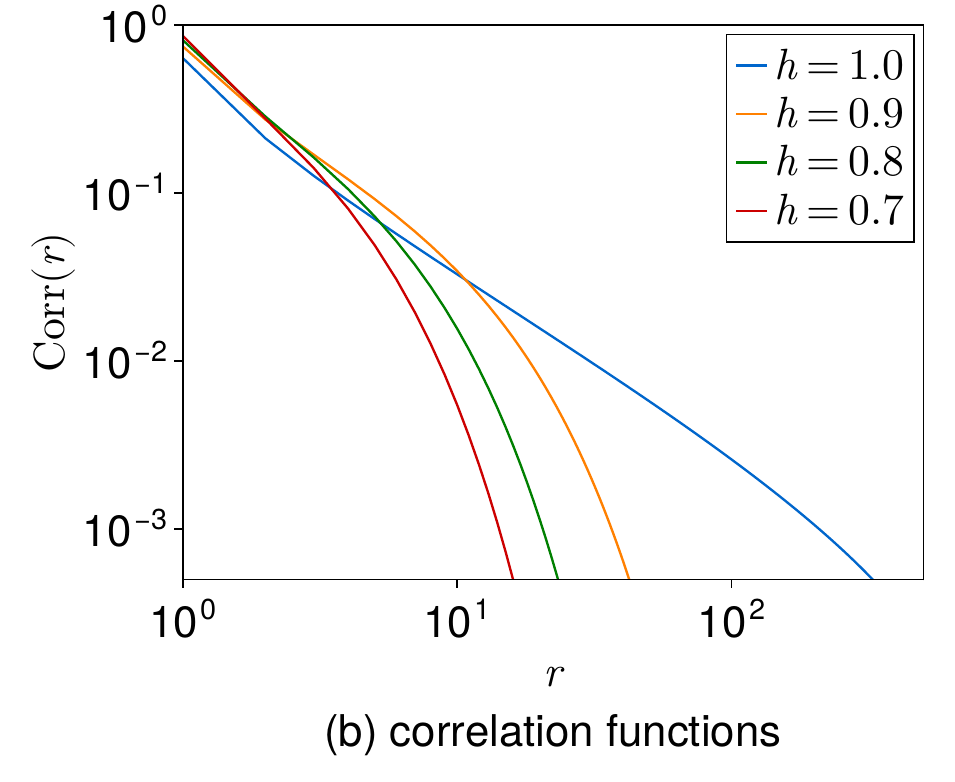}%
    \includegraphics[width=0.33\textwidth]{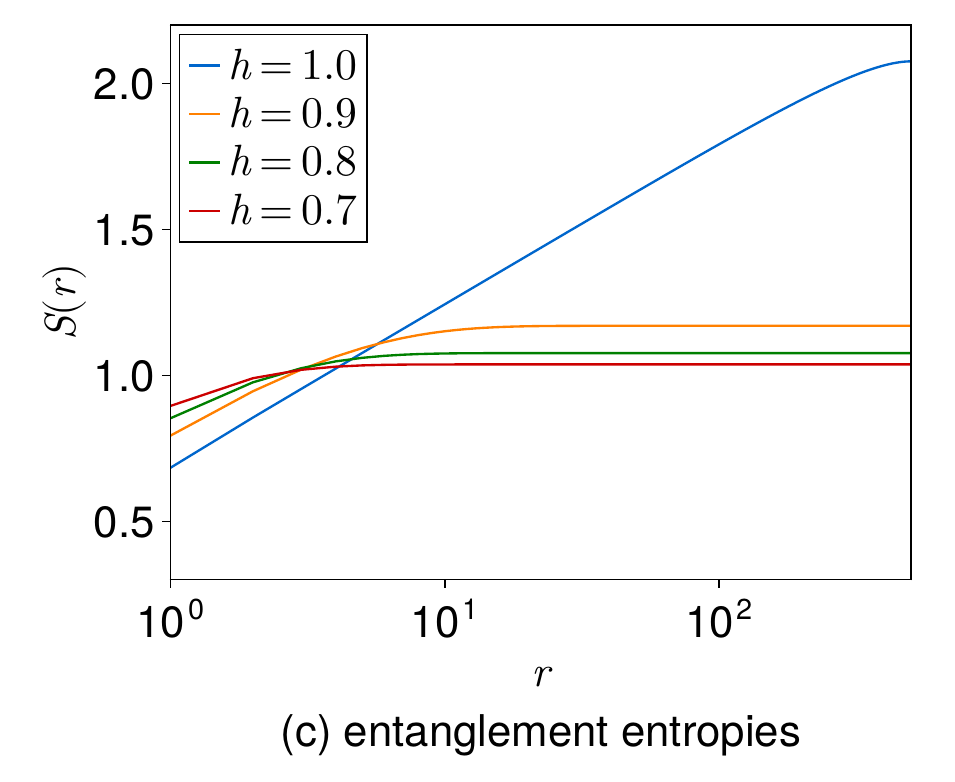}
    \includegraphics[width=0.33\textwidth]{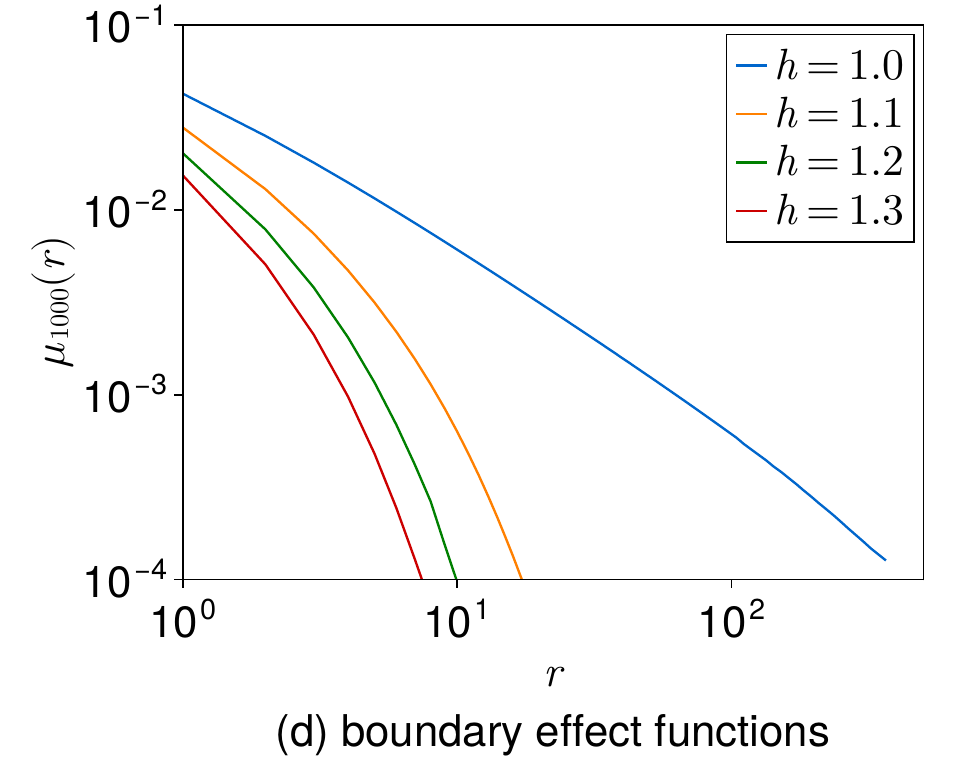}%
    \includegraphics[width=0.33\textwidth]{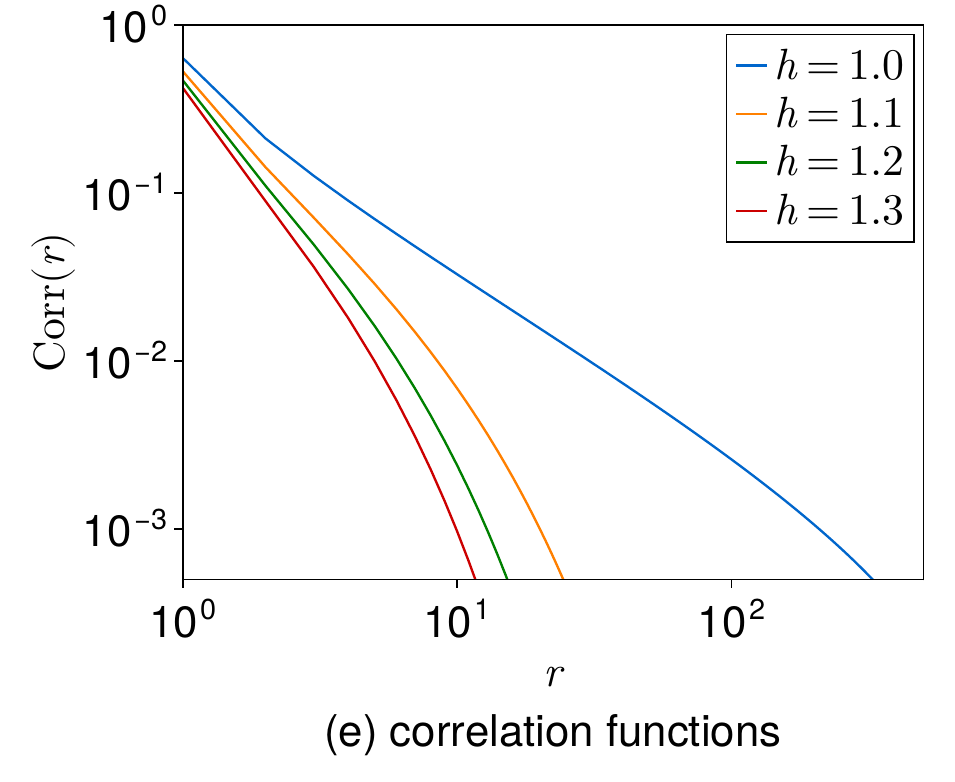}%
    \includegraphics[width=0.33\textwidth]{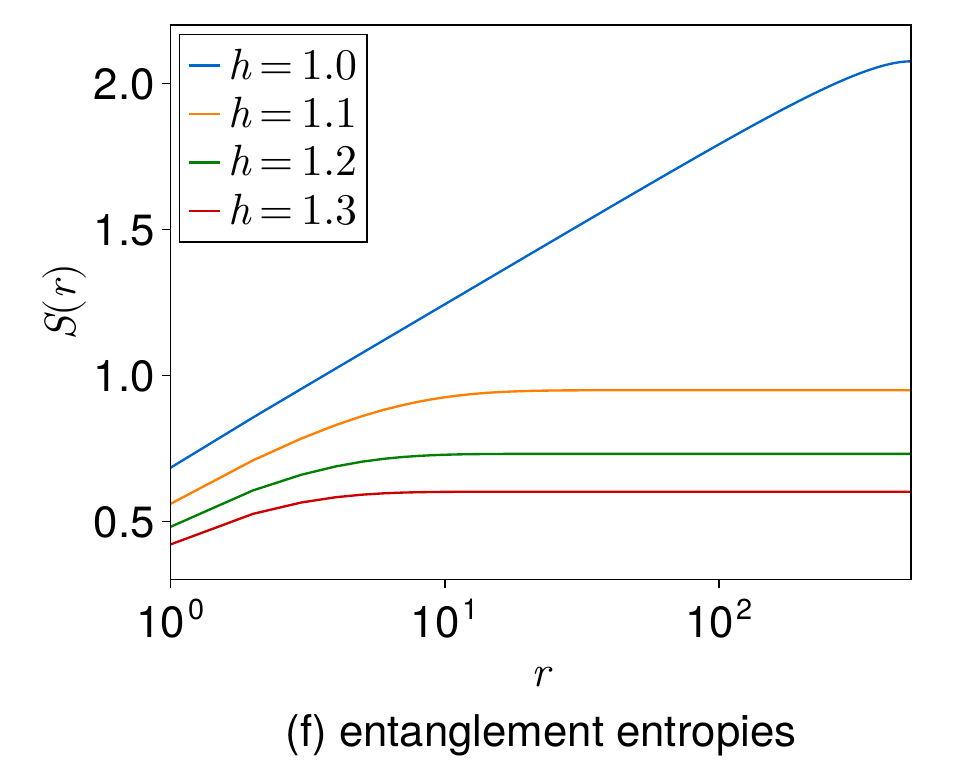}
    \caption{Boundary effect functions, correlation functions, and entanglement entropies for the transverse-field Ising model (Eq.~\eqref{eq:ising}) with varying $h$, while $n=1000$ fixed. The correlation function $\text{Corr}(r)$ indicates the correlation between the Majorana operators at site $n/4$ and $n/4+r$, defined as $\text{Corr}(r) = |\Gamma_{n/2,n/2+2r-1}|$. The entanglement entropy $S(r)$ represents the von Neumann entropy of the subchain of length $r$ from site $n/2+1$ to $n/2+r$.}
    \label{fig1}
\end{figure}

In Figs.~\ref{fig1}(a) and \ref{fig1}(d), we numerically obtain the BEF $\mu_{n}(r)$ for $n = 1000$ with varying parameter $h$.
The results indicate that as $h$ approaches the critical point from either side, the extent of the BEF increases and it eventually exhibits the behavior of polynomial decay.
This behavior is similar to that of the correlation function in the bulk.
For comparison, we plot in Figs.~\ref{fig1}(b) and \ref{fig1}(e) the correlation function $\Gamma_{n/2,n/2+2r-1}$ between sites $n/4$ and $n/4+r$.
Another important property of correlation is the scaling of the entanglement entropy.
We plot in Figs.~\ref{fig1}(c) and \ref{fig1}(f) the von Neumann entropy $S(r)$ of the subchain of length $r$ from site $n/2+1$ to $n/2+r$.
The logarithm growth of $S(r)$ at the critical point contrasts with its finite saturation in different parametric regimes.

\begin{figure}
    \includegraphics[width=0.4\textwidth]{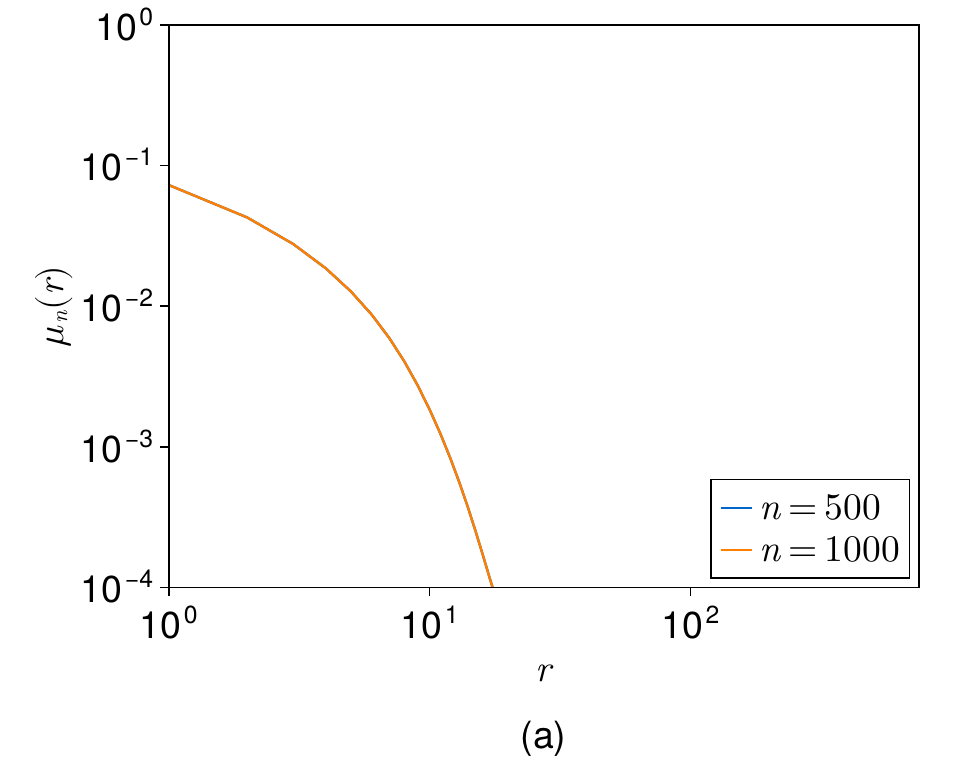}
    \includegraphics[width=0.4\textwidth]{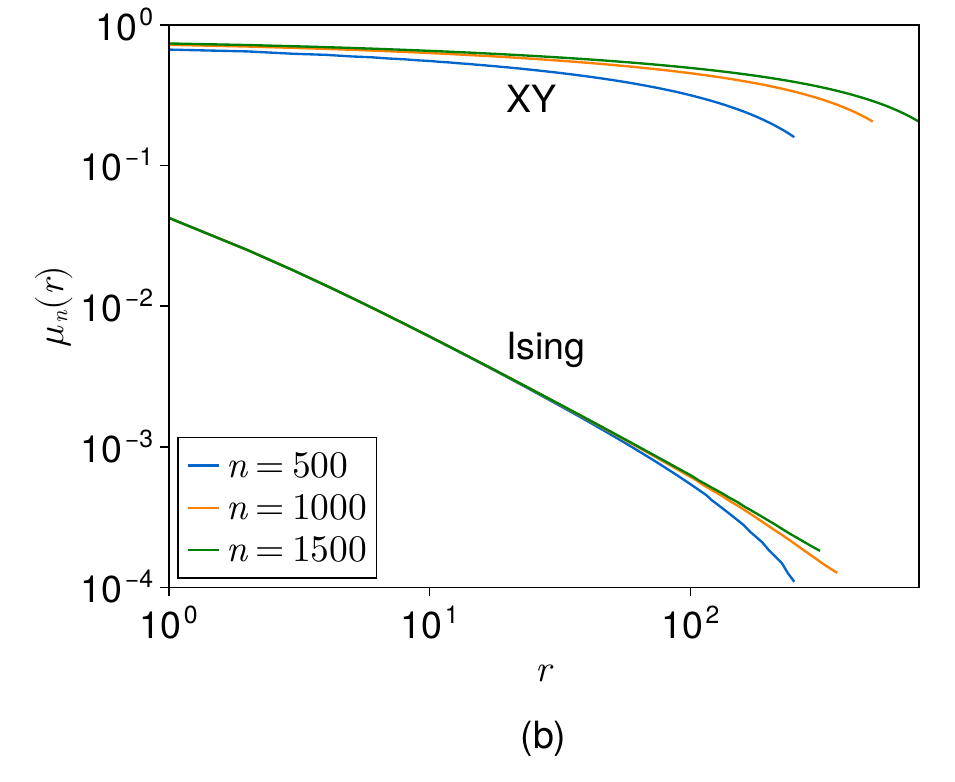}
    \caption{Typical finite-size scaling behavior of the BEF for (a) gapped and (b) gapless systems. (a) BEFs for the transverse-field Ising model (Eq.~\eqref{eq:ising}) with $h=0.8$. The two curves exactly overlap and are indistinguishable. (b) BEFs for the Ising model with $h=1$ and the $XY$ model (Eq.~\eqref{eq:xy}) with $h=0$.}
    \label{fig2}
\end{figure}

To check whether the observed finite length scale of the BEF is a finite-size effect or an intrinsic property, we present in Fig.~\ref{fig2}(a) the BEF for different system sizes $n$ while fixing $h=0.8$.
The results show that the BEFs $\mu_{500}(r)$ and $\mu_{1000}(r)$ for different system sizes completely overlap, indicating that the exponential decay is an intrinsic property.

\subsection{$XY$ model}

We also examine the BEF for the XY model, governed by the Hamiltonian
\begin{equation}
    H = \sum_{j=1}^{n-1} (\sigma_j^x \sigma_{j+1}^x + \sigma_j^y \sigma_{j+1}^y) + h \sum_{j=1}^{n} \sigma_j^z.
    \label{eq:xy}
\end{equation}
This model is thermodynamically gapless, resulting in a ground state exhibiting an infinite correlation length.
In Fig.~\ref{fig2}(b), we plot the BEF with increasing the system size $n$ while fixing $h=0$.
For comparison, we also present the analogous results for the Ising model at the critical point $h=1$.
Here, the result around $r=n/2$ should be interpreted carefully as the effect of the opposite boundary (the left hand side of the chain) coexists and interferes in the middle.
We thus obtain $\mu_n(r)$ only for $r < n/2$.
The result indicates that the boundary effect easily overwhelms the entire region for any system sizes within our computational capacity.
This implies that for only moderately large system sizes, the ground state properties are expected to deviate significantly from those in the thermodynamic limit.

\section{Conclusion}

In this work, we have derived the formula for the fidelity between two fermionic Gaussian states and applied the formula to obtain BEFs for standard free-fermion models.
While the BEF is known to upper-bound correlation functions in the bulk of the ground state, determining the tightness of the upper bound remains as a significant question.
If the tightness universally holds, the scaling of entanglement in ground states is proven to be strongly tied with the correlation length, thereby significantly advancing our theoretical foundation in diverse fields~\cite{eis10}.
The systems we numerically examined support this idea.
Further investigations in this direction could potentially provide new insights into the quantum information approaches to many-body theory.

\section*{Acknowledgement}

This work was supported by the National Research Foundation (NRF) of Korea under Grant No.~NRF-2022R1A4A1030660.

\appendix

\section{Derivation of the fidelity for fermionic Gaussian states}

The derivation here is based on Refs.~\cite{bra05,swi19}.
Let $\{c_j\}$ be the basis that block-diagonalizes the state $\rho$, hence
\begin{equation}
    \rho = \frac{1}{2^n} \prod_{j=1}^{n} (I + i \lambda_j c_{2j-1} c_{2j}).
    \label{eq:rho}
\end{equation}
Also, consider another state written, without loss of generality, as
\begin{equation}
    \sigma = \frac{1}{2^n} \prod_{j=1}^{n} (I + i \mu_j \tilde{c}_{2j-1} \tilde{c}_{2j}),
    \label{eq:sigma}
\end{equation}
where $\pmb{\tilde{c}} = \pmb{R}^T \pmb{c}$ with real orthogonal $\pmb{R}$.
The correlation matrices of these states are
\begin{align}
    \pmb{\Gamma}_\rho &\equiv 
    \bigoplus_{j=1}^{n} \left(
        \begin{matrix}0 & \lambda_j \\ -\lambda_j & 0 \end{matrix}
    \right), \label{eq:gamma_rho}\\
    \pmb{\Gamma}_\sigma &\equiv  
    \pmb{R} \left[\bigoplus_{j=1}^{n} 
    \left(\begin{matrix}0 & \mu_j \\ -\mu_j & 0 \end{matrix}\right) \right] \pmb{R}^T.
\end{align}

Our aim is to obtain the fidelity 
\begin{equation}
    F(\rho, \sigma) = \text{Tr} \sqrt{\sqrt{\rho} \sigma \sqrt{\rho}}.
\end{equation}
For this, we need mappings $\rho \mapsto \sqrt{\rho}$ and $(\rho,\sigma) \mapsto \sqrt{\rho} \sigma \sqrt{\rho}$.
Both the maps transform fermionic Gaussian states to fermionic Gaussian states.
They can thus be regarded as transformations of correlation matrices.

The mapping $\rho \mapsto \sqrt{\rho}$ is straightforward.
From the expression as in Eq.~\eqref{eq:ground}, one finds
\begin{equation}
    \begin{split}
        \sqrt{\rho} 
        &= \prod_{j=1}^{n} \left(\sqrt{\frac{1 - \lambda_j}{2}} f_j f_j^\dagger + \sqrt{\frac{1 + \lambda_j}{2}} f_j^\dagger f_j\right) \\
        &= \frac{1}{2^n} \prod_{j=1}^{n} \left(
            \frac{\sqrt{1 + \lambda_j} + \sqrt{1 - \lambda_j}}{\sqrt{2}} I 
            + i \frac{\sqrt{1 + \lambda_j} - \sqrt{1 - \lambda_j}}{\sqrt{2}} c_{2j-1} c_{2j}
        \right).
    \end{split}
    \label{eq:sqrt}
\end{equation}
The other mapping $(\rho,\sigma) \mapsto \sqrt{\rho} \sigma \sqrt{\rho}$ is, however, nontrivial to handle as Eqs.~\eqref{eq:rho} and \eqref{eq:sigma} are diagonalized in different bases, making $\sqrt{\rho} \sigma \sqrt{\rho}$ a complicated polynomial of the Majorana operators.
To deal with this problem, we need several tools.

\subsection{Choi-Jamiolkowski isomorphism}

Let us denote by $\mathcal{E}(\sigma) = \sqrt{\rho} \sigma \sqrt{\rho}$, which is a completely-positive map.
The Choi-Jamiolkowski isomorphism~\cite{cho75} works by introducing an (unnormalized) entangled state $\phi_{AB}$ of systems $A$ and $B$, and applying the input density operator $\sigma$ and the map $\mathcal{E}$, respectively, on systems $A$ and $B$ as $(\sigma_A \otimes \mathcal{E}_B)(\phi_{AB})$.
The state of system $B$, i.e., $\text{Tr}_A [ (\sigma_A \otimes \mathcal{E}_B)(\phi_{AB}) ]$, is then left as the desired output of the map $\mathcal{E} (\sigma)$.
This procedure is equivalent to that of quantum teleportation~\cite{nie11}.
In this way, the above-mentioned difficulty is reduced to the problem of handling the trace at the last step, which is somehow manageable.

We need to tailor the Choi-Jamiolkowski isomorphism for the Majorana algebra.
Consider an unnormalized state
\begin{equation}
    \begin{split}
        \phi_{AB} 
        &\equiv \frac{1}{2^n} \prod_{j=1}^{2n} (I_A \otimes I_B + i c_j^A \otimes c_j^B) \\
        &= \frac{1}{2^n} \left( I_A \otimes I_B + \sum_{m=1}^{n} \sum_{j_1 < \cdots < j_{2m}} (-1)^m c_{j_1}^A  \cdots c_{j_{2m}}^A \otimes c_{j_1}^B  \cdots c_{j_{2m}}^B  \right) + ...,
    \end{split}
\end{equation}
where the last dots abbreviate the remaining odd-power terms.
In our formalism, only even powers of Majorana operators are used.
Here, the two systems $A$ and $B$ are completely independent, which means that $[c_j^A, c_k^B] = 0$.
Thanks to the linearity of the map, we have
\begin{equation}
    \begin{split}
        &(I_A \otimes \mathcal{E}_B)(\phi_{AB}) \\
        &= \frac{1}{2^n} \left[ 
            I_A \otimes \mathcal{E}_B (I_B) + 
            \sum_{m=1}^{n} 
            \sum_{j_1 < \cdots < j_{2m}} 
            (-1)^m c_{j_1}^A \cdots c_{j_{2m}}^A \otimes 
            \mathcal{E}_B \left(  
                c_{j_1}^B \cdots c_{j_{2m}}^B 
                \right)  
        \right] + \cdots.
    \end{split}
    \label{eq:phi_AB}
\end{equation}
The last step is to plug the state $\sigma$ into system $A$ and trace it.
Note that as $c_j^2 = I$, any polynomial of Majorana operators, as well as density operators, can be written \emph{uniquely} as
\begin{equation}
    \sigma = \frac{1}{2^n}\left(\alpha_0 I + \sum_{n=1}^{2m} \sum_{1\le j_1 < j_2 < \cdots < j_m \le 2n} \alpha_{j_1 j_2 \cdots j_m} c_{j_1} c_{j_2} \cdots c_{j_m}\right)
    \label{eq:unique}
\end{equation}
with complex coefficients $\alpha_0$ and $\alpha_{j_1 j_2 \cdots j_m}$.
Using the property in Eq.~\eqref{eq:trace}, the coefficients are found to be $\alpha_0 = \text{Tr}(\sigma)$ and $\alpha_{j_1 \cdots j_{2m}} = (-1)^m \text{Tr}(\sigma c_{j_1} \cdots c_{j_{2m}})$.
Using the same property, we end up with
\begin{equation}
    \begin{split}
        \text{Tr}_A \left[
            (\sigma_A \otimes \mathcal{E}_B) (\phi_{AB})
        \right]
        &= \frac{1}{2^n} \left[ 
            \alpha_0 \mathcal{E}_B (I_B) + \sum_{m=1}^{n} \sum_{j_1 < \cdots < j_{2m}}
            \alpha_{j_1 \cdots j_{2m}}
            \mathcal{E}_B \left( c_{j_1}^B \cdots c_{j_{2m}}^B \right)
        \right] \\
        &= \mathcal{E}_{A \rightarrow B}(\sigma_A).
    \end{split}
    \label{eq:map}
\end{equation}

Coming back to our problem, we perform the above task for $\mathcal{E}(\sigma) = \sqrt{\rho} \sigma \sqrt{\rho}$.
The operator~\eqref{eq:phi_AB} can be obtained without difficulty because, first, we can use the same basis as in Eq.~\eqref{eq:rho} throughout and, second, all terms are even powers of Majorana operators and hence all commute unless they commonly contain the same Majorana operator.
Using Eq.~\eqref{eq:sqrt}, we obtain
\begin{equation}
    \begin{split}
        (I_A \otimes \mathcal{E}_B)(\phi_{AB})
        &= \frac{1}{2^{2n}} \prod_{j=1}^{n} \left[ 
            I_A \otimes I_B + i \lambda_j 
            \left( c_{2j-1}^A c_{2j}^A \otimes I_B 
            + I_A \otimes c_{2j-1}^B c_{2j}^B \right) \right.\\
        & + i \sqrt{1 - \lambda_j^2} 
        \left( c_{2j-1}^{A} \otimes c_{2j-1}^{B} 
        + c_{2j}^{A} \otimes c_{2j}^{B}  \right)
        - \left. c_{2j-1}^{A} c_{2j}^{A} \otimes 
        c_{2j-1}^{B} c_{2j}^{B} \right].
    \end{split}
    \label{eq:iephi}
\end{equation}

The next step is to apply the density operator~\eqref{eq:sigma} to operator~\eqref{eq:iephi} and trace out system $A$, as in Eq.~\eqref{eq:map}.
For this, we need another important tool, namely, the Grassmann representation.

\subsection{Grassmann representation}

Consider $2n$ Grassmann numbers $\theta_j$, which satisfy the anticommutation relation $\{\theta_j, \theta_k\} = 0$.
The important feature of the Grassmann number is that the derivative and integral is defined identically as
\begin{equation}
    \int d\theta_j \, \theta_k = \frac{\partial \theta_k}{\partial \theta_j} = \delta_{jk}.
    \label{eq:int1}
\end{equation}
Let $\pmb{\theta} \equiv (\theta_1 ~ \cdots ~ \theta_{2n})^T$ be the column vector of the Grassmann numbers.
We define the multiple integration over the Grassmann numbers as
\begin{equation}
    \int d\pmb{\theta} \equiv \int d\theta_{2n} d\theta_{2n-1} \cdots d\theta_{1},   
    \label{eq:int2}
\end{equation}
which implies $\int d\pmb{\theta} \, \theta_1 \theta_2 \cdots \theta_{2n} = 1$.

The Grassmann representation $A^{\pmb{\theta}}$ of polynomial $A(\pmb{c})$ of Majorana operators is obtained by replacing every $c_j$ with $\theta_j$~\footnote{Before the replacement, the function of Majorana operators should be converted into the unique form of Eq.~\eqref{eq:unique} to avoid ambiguity. For example, the Grassmann representation of $c_1 c_2 c_1$ is $-\theta_2$, not $\theta_1 \theta_2 \theta_1 = 0$.}.
Thanks to the property $\theta_j^2 = 0$, the Grassmann representation of the density matrix~\eqref{eq:rho} is given by the correlation matrix~\eqref{eq:gamma_rho} as
\begin{equation}
    \begin{split}
    \rho^{\pmb{\theta}} 
    &= \frac{1}{2^n} \prod_{j=1}^{n} 
    \left( I + i \lambda_j {\theta}_{2j-1} 
    {\theta}_{2j} \right) 
    = \frac{1}{2^n} \prod_{j=1}^{n} \exp 
    \left(i \lambda_j {\theta}_{2j-1} 
    {\theta}_{2j}\right) \\
    &= \frac{1}{2^n} 
    \exp\left( i \sum_{j=1}^{n} \lambda_j 
    {\theta}_{2j-1} {\theta}_{2j} \right) 
    = \frac{1}{2^n} \exp\left[ \frac{i}{2} 
    \pmb{{\theta}}^T \left\{ \bigoplus_{j=1}^{n} 
    \left(\begin{matrix} 0 & \lambda_j \\ - \lambda_j & 0 \end{matrix} \right) \right\} 
    \pmb{{\theta}} \right] \\
    &= \frac{1}{2^n} \exp \left( \frac{i}{2} \pmb{\theta}^T 
    \pmb{\Gamma}_\rho \pmb{\theta}  \right).
    \end{split}
    \label{eq:grassmann}
\end{equation}
Similarly, $\sigma^{\pmb{\theta}} = \frac{1}{2^n} \exp \left( \frac{i}{2} \pmb{\theta}^T 
\pmb{\Gamma}_\sigma \pmb{\theta}  \right)$.
Unlike Eq.~\eqref{eq:thermal}, this representation directly contains the correlation matrix in the exponent, eliminating the issue of infinity when dealing with ground states.
Furthermore, the exponential form greatly facilitates the algebra in contrast to Eq.~\eqref{eq:ground}.

For later use, let us derive two important identities.
First, let $\pmb{M}=-\pmb{M}^T$ be any real skew-symmetric matrix.
The following Gaussian integration is very useful:
\begin{equation}
    \begin{split}
        &\int D\pmb{\theta} \exp \left( 
            \frac{i}{2} \pmb{\theta}^T \pmb{M} \pmb{\theta}
        \right) \\
        &=  \frac{1}{n!} \left(\frac{i}{2}\right)^n
        \int D\pmb{\theta} \sum_{\sigma \in S_{2n}}
        M_{\sigma(1), \sigma(2)} \cdots 
        M_{\sigma(2n-1), \sigma(2n)}
        \theta_{\sigma(1)} \cdots
        \theta_{\sigma(2n)} \\
        &= \frac{1}{n!} \left(\frac{i}{2}\right)^n 
        \int D\pmb{\theta} 
        \sum_{\sigma \in S_{2n}} 
        M_{\sigma(1),\sigma(2)} \cdots 
        M_{\sigma(2n-1),\sigma(2n)}
        (-1)^\sigma
        \theta_{1} \cdots
        \theta_{2n} \\
        &= i^n \text{Pf} (\pmb{M}),
    \end{split}
    \label{eq:integral}
\end{equation}
where
\begin{equation}
    \text{Pf} (\pmb{M}) 
\equiv \frac{1}{2^n n!} \sum_{\sigma \in S_{2n}} 
(-1)^\sigma M_{\sigma(1),\sigma(2)} \cdots M_{\sigma(2n-1),\sigma(2n)}
= \sqrt{\det(\pmb{M})}
\end{equation}
is the Pfaffian of matrix $\pmb{M}$ and $S_{2n}$ denotes the symmetric group of degree $2n$.
In the first line of Eq.~\eqref{eq:integral}, only the highest-order terms in the expansion survive due to the integration defined as in Eqs.~\eqref{eq:int1} and \eqref{eq:int2}.

Another useful identity is
\begin{equation}
    \text{Tr} (AB) 
    = (-2)^n \int D \pmb{\theta} D \pmb{\eta} \exp \left(
        \pmb{\eta}^T \pmb{\theta}
    \right)
    A^{\pmb{\eta}} B^{\pmb{\theta}},
    \label{eq:traceAB}
\end{equation}
where $\pmb{\eta} = (\eta_1 ~ \cdots ~ \eta_{2n})^T$ is also a vector of Grassmann numbers.
To see this, express $A$ and $B$ as in Eq.~\eqref{eq:unique}: $A = \alpha_0 I + \sum \alpha_{j_1 \cdots j_m} c_{j_1} \cdots c_{j_m}$ and $B = \beta_0 I + \sum \beta_{j_1 \cdots j_m} c_{j_1} \cdots c_{j_m}$.
From the property in Eq.~\eqref{eq:trace}, we find $\text{Tr} (AB) = \alpha_0 \beta_0 \text{Tr}(I) + \sum \alpha_{j_1 \cdots j_m} \beta_{j_1 \cdots j_m} \text{Tr} (c_{j_1} \cdots c_{j_m}c_{j_1} \cdots c_{j_m}) = 2^n \{\alpha_0 \beta_0 + \sum (-1)^{\frac{m(m-1)}{2}} \alpha_{j_1 \cdots j_m} \beta_{j_1 \cdots j_m} \}$.
To see that this is identical to the right hand side in Eq.~\eqref{eq:traceAB}, note that 
\begin{equation}
    \exp \left(\pmb{\eta}^T \pmb{\theta}\right) 
    = \prod_{j=1}^{2n} \left( I + \eta_{j} \theta_{j}\right)
    = I + \sum_{m=1}^{2n} \sum_{j_1 < \cdots < j_m} 
    \eta_{j_1} \theta_{j_1} \cdots \eta_{j_m} \theta_{j_m},
    \label{eq:etatheta}
\end{equation}
where $[\eta_j \theta_j, \eta_k \theta_k] = 0$ is used.
As $\eta_j$ and $\theta_j$ appear in pairs, every non-vanishing term in Eq.~\eqref{eq:traceAB} contains the product of identical monomials from $A^{\pmb{\eta}}$ and $B^{\pmb{\theta}}$, and a term from Eq.~\eqref{eq:etatheta}, making a highest-order term in total.
It is thus sufficient to check if the identity holds for $A^{\pmb{\eta}} = \eta_{j_1} \cdots \eta_{j_m}$ and $B^{\pmb{\theta}} = \theta_{j_1} \cdots \theta_{j_m}$.
Note that $A^{\pmb{\eta}} B^{\pmb{\theta}} = (-1)^{\frac{m(m-1)}{2}} \eta_{j_1}\theta_{j_1} \cdots \eta_{j_m}\theta_{j_m}$.
As $[\eta_j\theta_j, \eta_k\theta_k]=0$, we find
\begin{equation}
    \begin{split}
        \int D \pmb{\theta} \, D \pmb{\eta} \, \exp \left(
            \pmb{\eta}^T \pmb{\theta}
        \right) A^{\pmb{\eta}} B^{\pmb{\theta}}
        &= (-1)^{\frac{m(m-1)}{2}} \int D \pmb{\theta} \, D \pmb{\eta} \, 
        \eta_1 \theta_1 \cdots \eta_{2n} \theta_{2n} \\
        &= (-1)^{\frac{m(m-1)}{2}} \int D \pmb{\theta} \, D \pmb{\eta} \, 
        (-1)^n \eta_1 \cdots \eta_{2n} \theta_1 \cdots \theta_{2n} \\
        &= (-1)^n (-1)^{\frac{m(m-1)}{2}},
    \end{split}
\end{equation}
confirming the identity~\eqref{eq:traceAB}.

\subsection{Fidelity}

We are now equipped with all the necessary tools.
The Grassmann representation of Eq.~\eqref{eq:iephi} becomes
\begin{equation}
    \begin{split}
        &(I_A \otimes \mathcal{E}_B)(\phi_{AB})^{\pmb{\theta}} \\
        &= \frac{1}{2^{2n}} \prod_{j=1}^{n} \exp 
        \left[ i \lambda_j \left( {\theta}_{2j-1}^{A}
        {\theta}_{2j}^{A} + {\theta}_{2j-1}^{B} 
        {\theta}_{2j}^{B}  \right) + i \sqrt{1 - \lambda_j^2} 
        \left( \theta_{2j-1}^{A} \theta_{2j-1}^{B} +
        \theta_{2j}^{A} \theta_{2j}^{B}  \right)    \right] \\
        &= \frac{1}{2^{2n}} \exp\left( \frac{i}{2} 
        \pmb{\theta}_A^T \pmb{\Gamma}_\rho \pmb{\theta}_A 
        + i \, \pmb{\theta}_A^T \! \sqrt{\pmb{I} + 
        \pmb{\Gamma}_\rho^2} \, \pmb{\theta}_B + \frac{i}{2} 
        \pmb{\theta}_B^T \pmb{\Gamma}_\rho \pmb{\theta}_B \right).
    \end{split}
\end{equation}
Note that in the first line, the second-order terms in the expansion do not vanish.
Here, we have used the independency of the two systems $A$ and $B$, implied by $[\theta_j^A, \theta_k^B] = 0$.
In the second line, we have used $\sqrt{\pmb{I} + \pmb{\Gamma}_\rho^2} = \bigoplus_{j=1}^{n}
\sqrt{1 - \lambda_j^2} \left(\begin{matrix} 1 & 0 \\ 0 & 1 \end{matrix}\right)$.
Using Eq.~\eqref{eq:traceAB},
\begin{equation}
    \begin{split}
        &\text{Tr}_A \left[ 
            (\sigma_A \otimes \mathcal{E}_B) (\phi_{AB}) 
        \right]^{\pmb{\theta}} \\
        &= (-2)^n \int D \pmb{\theta}_A D \pmb{\eta}_A
        \exp \left( \pmb{\eta}_A^T \pmb{\theta}_A \right) 
        \sigma_A^{\pmb{\eta}} \left[ 
            (I_A \otimes \mathcal{E}_B) (\phi_{AB}) 
        \right]^{\pmb{\theta}} \\
        &= \left(- \frac{1}{2^2}\right)^{n} 
        \int D \pmb{\theta}_A D \pmb{\eta}_A
        \exp \left( 
            \frac{i}{2} \pmb{\eta}_A^T \pmb{\Gamma}_\sigma 
            \pmb{\eta}_A  
            + \pmb{\eta}_A^T \pmb{\theta}_A \right. \\  
        & \hspace{14em} \left. + \frac{i}{2} \pmb{\theta}_A^T \pmb{\Gamma}_\rho 
            \pmb{\theta}_A 
            + i \, \pmb{\theta}_A^T \! 
            \sqrt{\pmb{I} + \pmb{\Gamma}_\rho^2} \, 
            \pmb{\theta}_B 
            + \frac{i}{2} \pmb{\theta}_B^T \pmb{\Gamma}_\rho 
            \pmb{\theta}_B 
        \right)  
    \end{split}
\end{equation}
The first part in the exponent can be rewritten as $\frac{i}{2} \pmb{\eta}_A^T \pmb{\Gamma}_\sigma \pmb{\eta}_A + \pmb{\eta_A^T} \pmb{\theta}_A = \frac{i}{2} \pmb{\eta}_A^T \pmb{\Gamma}_\sigma \pmb{\eta}_A + \frac{1}{2}\pmb{\eta_A^T} \pmb{\theta}_A - \frac{1}{2}\pmb{\theta_A^T} \pmb{\eta}_A = \frac{i}{2} ( \pmb{\eta}_A^T +i \pmb{\theta}_A^T \pmb{\Gamma}_\sigma^{-1} ) \pmb{\Gamma}_\sigma ( \pmb{\eta}_A -i \pmb{\Gamma}_\sigma^{-1} \pmb{\theta}_A ) - \frac{i}{2} \pmb{\theta}_A^T \pmb{\Gamma}_\sigma^{-1} \pmb{\theta}_A$ for the Gaussian integration.
Note that $\left\{\eta_j^A, \theta_k^A\right\} = 0$.
From Eq.~\eqref{eq:integral}, we obtain
\begin{equation}
    \begin{split}
        &\text{Tr}_A \left[ 
            (\sigma_A \otimes \mathcal{E}_B) (\phi_{AB}) 
        \right]^{\pmb{\theta}} \\
        &= \left(- \frac{i}{2^2}\right)^{n} 
        \text{Pf} \left(\pmb{\Gamma}_\sigma \right)
        \int D \pmb{\theta}_A \exp \left[ 
            \frac{i}{2} \pmb{\theta}_A^T \left(
                \pmb{\Gamma}_\rho - \pmb{\Gamma}_\sigma^{-1}
            \right) \pmb{\theta}_A 
            + i \, \pmb{\theta}_A^T \! \sqrt{\pmb{I} + 
            \pmb{\Gamma}_\rho^2} \, \pmb{\theta}_B + \frac{i}{2} 
            \pmb{\theta}_B^T \pmb{\Gamma}_\rho \pmb{\theta}_B 
        \right]            
    \end{split}
\end{equation}
Let us denote $\pmb{A} \equiv \pmb{\Gamma}_\rho - \pmb{\Gamma}_\sigma^{-1}$ and $\pmb{B} \equiv \sqrt{ \pmb{I} + \pmb{\Gamma}_\rho^2}$ to rearrange the exponent as $\frac{i}{2} \pmb{\theta}_A^{T} \pmb{A} \pmb{\theta}_A + i \pmb{\theta}_A^T \pmb{B} \pmb{\theta}_B = \frac{i}{2} \pmb{\theta}_A^{T} \pmb{A} \pmb{\theta}_A + \frac{i}{2} \pmb{\theta}_A^T \pmb{B} \pmb{\theta}_B + \frac{i}{2} \pmb{\theta}_B^T \pmb{B} \pmb{\theta}_A = \frac{i}{2} ( \pmb{\theta}_A^T + \pmb{\theta}_B^T \pmb{B} \pmb{A}^{-1} ) \pmb{A} ( \pmb{\theta}_A + \pmb{A}^{-1} \pmb{B} \pmb{\theta}_B ) - \frac{i}{2} \pmb{\theta}_B^T \pmb{B} \pmb{A}^{-1} \pmb{B} \pmb{\theta}_B$.
Using Eq.~\eqref{eq:integral} again, we obtain
\begin{equation}
    \begin{split}
        &\text{Tr}_A \left[ 
            (\sigma_A \otimes \mathcal{E}_B) (\phi_{AB})
        \right]^{\pmb{\theta}} \\
        &= \frac{1}{2^{2n}} \text{Pf} \left(
            \pmb{\Gamma}_\sigma 
        \right) \text{Pf} \left( 
            \pmb{\Gamma}_\rho - \pmb{\Gamma}_\sigma^{-1}
        \right)  \exp \left[ 
            \frac{i}{2} \pmb{\theta}^T \left\{
                \pmb{\Gamma}_\rho 
                -\sqrt{\pmb{I} + \pmb{\Gamma}_\rho^2} \left(
                    \pmb{\Gamma}_\rho - \pmb{\Gamma}_\sigma^{-1}
                \right)^{-1} \!\! 
                \sqrt{\pmb{I} + \pmb{\Gamma}_\rho^2}
            \right\} \pmb{\theta} 
        \right] \\
        &= \left[
            \det\left(
                \frac{\pmb{\Gamma}_\rho \pmb{\Gamma}_\sigma - \pmb{I}}{2}
            \right)
        \right]^{\frac{1}{2}} \frac{1}{2^{n}} \exp \left[ 
            \frac{i}{2} \pmb{\theta}^T 
            \left\{
                \pmb{\Gamma}_\rho 
                - \sqrt{\pmb{I} + \pmb{\Gamma}_\rho^2}  
                \left(
                    \pmb{\Gamma}_\rho - \pmb{\Gamma}_\sigma^{-1}
                \right)^{-1} \!\!
                \sqrt{\pmb{I} + \pmb{\Gamma}_\rho^2} 
            \right\} \pmb{\theta}
        \right] \\
        &= \left(
            \sqrt{\rho} \sigma \sqrt{\rho}
        \right)^{\, \pmb{\theta}}
    \end{split}
    \label{eq:rsr}
\end{equation}
This representation has the same form as Eq.~\eqref{eq:grassmann} up to the global factor, where
\begin{equation}
    \pmb{\Lambda} \equiv 
    \pmb{\Gamma}_\rho
    - \sqrt{\pmb{I} + \pmb{\Gamma}_\rho^2} 
    \left( 
        \pmb{\Gamma}_\rho - \pmb{\Gamma}_\sigma^{-1} 
    \right)^{-1} \! \! \sqrt{\pmb{I} + \pmb{\Gamma}_\rho^2} = - \pmb{\Lambda}^T
\end{equation}
plays the role of the correlation matrix.
This implies that we have obtained the Majorana polynomial of $\sqrt{\rho} \sigma \sqrt{\rho}$.
We can thus obtain the eigenvalues of $\sqrt{\sqrt{\rho} \sigma \sqrt{\rho}}$ by treating the polynomial as in Eq.~\eqref{eq:sqrt}.
Block diagonalizing $\pmb{\Lambda}$ as
\begin{equation}
    \pmb{\Lambda} = \pmb{W} 
    \left[ \bigoplus_{j=1}^{n} 
    \left(\begin{matrix} 0 & \nu_j \\ - \nu_j & 0 \end{matrix}\right) \right] \pmb{W}^T,
\end{equation}
we obtain
\begin{equation}
    \begin{split}
        F(\rho, \sigma) 
        &= \left[
            \det\left(
                \frac{\pmb{\Gamma}_\rho \pmb{\Gamma}_\sigma - \pmb{I}}{2}
            \right)
        \right]^{\frac{1}{4}}
        \prod_{j=1}^{n} \left( 
            \sqrt{\frac{1 - \nu_j}{2}} + \sqrt{\frac{1 + \nu_j}{2}}
        \right) \\
        &= \left[
            \det\left(
                \frac{\pmb{\Gamma}_\rho \pmb{\Gamma}_\sigma - \pmb{I}}{2}
            \right)
        \right]^{\frac{1}{4}}
        \prod_{j=1}^{n} 
        \sqrt{1 + \sqrt{1 - \nu_j^2}} \\
        &= \left[
            \det\left(
                \frac{\pmb{\Gamma}_\rho \pmb{\Gamma}_\sigma - \pmb{I}}{2}
            \right)
        \right]^{\frac{1}{4}}
        \left[            
            \det \left(
                \pmb{I} + \sqrt{\pmb{I} + \pmb{\Lambda}^2}
            \right)
        \right]^{\frac{1}{4}}.
    \end{split}
    \label{eq:f2}
\end{equation}
We proceed further to handle the potential singularity in matrix $\pmb{\Lambda}$ properly.
Using $(\pmb{\Gamma}_\rho - \pmb{\Gamma}_\sigma^{-1})^{-1} = [ ( \pmb{\Gamma}_\rho \pmb{\Gamma}_\sigma - \pmb{I} ) \pmb{\Gamma}_\sigma^{-1} ]^{-1} = \pmb{\Gamma}_\sigma ( \pmb{\Gamma}_\rho \pmb{\Gamma}_\sigma - \pmb{I} )^{-1}$, we can rewrite
\begin{equation}
    \begin{split}
        \pmb{\Lambda}
        &= \left( 
            \pmb{I} + \pmb{\Gamma}_\rho^{2}
        \right)^{1/2} \left[
            \left( 
                \pmb{I} + \pmb{\Gamma}_\rho^{2}
            \right)^{-1} \pmb{\Gamma}_\rho 
            - \pmb{\Gamma}_\sigma \left(
                \pmb{\Gamma}_\rho \pmb{\Gamma}_\sigma - \pmb{I}
            \right)^{-1}
        \right] \left( 
            \pmb{I} + \pmb{\Gamma}_\rho^{2}
        \right)^{1/2} \\
        &= \left( 
            \pmb{I} + \pmb{\Gamma}_\rho^{2}
        \right)^{-1/2} \left[
            \pmb{\Gamma}_\rho 
            - \left(
                \pmb{I} + \pmb{\Gamma}_\rho^2 
            \right) \pmb{\Gamma}_\sigma \left(
                \pmb{\Gamma}_\rho \pmb{\Gamma}_\sigma - \pmb{I}
            \right)^{-1}
        \right] \left( 
            \pmb{I} + \pmb{\Gamma}_\rho^{2}
        \right)^{1/2} \\
        &= \left( 
            \pmb{I} + \pmb{\Gamma}_\rho^{2}
        \right)^{-1/2} \left[
            \pmb{\Gamma}_\rho \left(
                \pmb{\Gamma}_\rho \pmb{\Gamma}_\sigma - \pmb{I}
            \right)
            - \left(
                \pmb{I} + \pmb{\Gamma}_\rho^2 
            \right) \pmb{\Gamma}_\sigma 
        \right] \left(
            \pmb{\Gamma}_\rho \pmb{\Gamma}_\sigma - \pmb{I}
        \right)^{-1} \left( 
            \pmb{I} + \pmb{\Gamma}_\rho^{2}
        \right)^{1/2} \\
        &= - \left( 
            \pmb{I} + \pmb{\Gamma}_\rho^{2}
        \right)^{-1/2} \left(
            \pmb{\Gamma}_\rho + \pmb{\Gamma}_\sigma
        \right) \left(
            \pmb{\Gamma}_\rho \pmb{\Gamma}_\sigma - \pmb{I}
        \right)^{-1} \left( 
            \pmb{I} + \pmb{\Gamma}_\rho^{2}
        \right)^{1/2}.
    \end{split},
\end{equation} 
Plugging this into Eq.~\eqref{eq:f2} and using the property of determinants, we end up with the fidelity~\eqref{eq:fidelity}.


%

\end{document}